\documentclass[10pt,letterpaper]{article}
\usepackage{opex3}\usepackage{hyperref}\pdfoutput=1

\usepackage[compress]{cite}
\usepackage{amssymb}
\usepackage{graphicx}

\newcommand\etal{et al.}

\newcommand{\Ga}{\alpha}

\newcommand{\Gd}{\delta}

\newcommand{\Gg}{\gamma}

\newcommand{\CO}{{\cal O}}

\def\Bb{{\bf b}}

\def\Bp{{\bf p}}

\def\Bx{{\bf x}}

\def\Bz{{\bf z}}
\def\BA{{\bf A}}
\def\BB{{\bf B}}

\def\BZ{{\bf Z}}

\def \om {\omega}
\def \la {\lambda}

\def \ba {\begin{array}}
\def \ea {\end{array}}

\newcommand{\abs}[1]{\left|{#1}\right|}

\newcommand{\ceil}[1]{\lceil{#1}\rceil}

\newcommand{\Bzero}{\mathbf{0}}

\newcommand{\Hg}{\widehat{g}}

\newcommand{\figref}[1]{Fig.~\ref{fig:#1}}

\begin{document}
\title{Broadband Exterior Cloaking}
\author{Fernando Guevara Vasquez, Graeme W. Milton, Daniel
Onofrei}
\address{Department of Mathematics, University of Utah, Salt Lake City
UT 84112, USA}
\email{fguevara@math.utah.edu}

\begin{abstract}
It is shown how a recently proposed method of cloaking is effective over
a broad range of frequencies. The method is based on three or more
active devices. The devices, while not radiating significantly, create a
``quiet zone'' between the devices where the wave amplitude is small.
Objects placed within this region are virtually invisible. The cloaking
is demonstrated by simulations with a broadband incident pulse. 
\end{abstract}

\ocis{(260.0260) Physical optics; (350.7420) Waves; (160.4760) Optical properties}

\bibliographystyle{osajnl}
\bibliography{sciref}
\pagestyle{plain} 

\section{Introduction}
A tremendous amount of interest and excitement has been generated by
recent strides towards making objects invisible, not by camouflage, but
by manipulating the fields in such a way that the cloaking device and
the object to be cloaked scatter very little radiation in any direction
and do not absorb it. Here using a new method of active exterior
cloaking, described in \cite{Vasquez:2009:AEC} for single frequency
waves,  we demonstrate how an object can be cloaked against an incoming
broadband pulse. To our knowledge this is the first example showing
cloaking of an object against an incoming pulse. It uses active cloaking
devices to generate anomalous localized waves which cancel the incident
waves within the cloaking region to create a ``quiet zone'', within
which objects can be hidden. Because active devices, rather than
materials, are used to generate the  anomalous localized waves, one may
superimpose the results for different frequencies to obtain broadband
cloaking. Our method requires one to know the form of the incoming pulse
in advance, since the fields generated by the cloaking devices are
tailored to the incoming fields.

Dolin \cite{Dolin:1961:PCT}, Kerker \cite{Kerker:1975:IB}, and Al\'u and
Engheta \cite{Alu:2005:ATP} realized that certain objects could be made
invisible by coating them with an appropriate material tailored
according to the object to be cloaked. A breakthrough came with the work
of Greenleaf \etal \cite{Greenleaf:2003:ACC}, for conductivity,
Leonhardt \cite{Leonhardt:2006:OCM}, for geometric optics, and Pendry
\etal \cite{Pendry:2006:CEM} for electromagnetism who showed that
materials could guide fields around a region, leaving a ``quiet'' zone
in that region within which objects could be placed without disturbing
the surrounding field. This idea was extended to acoustics
\cite{Chen:2007:ACT,Greenleaf:2007:FWI,Cummer:2008:STD,Norris:2008:ACT},
elastodynamics \cite{Milton:2006:CEM}, and water waves
\cite{Farhat:2008:BCA}, and has been confirmed experimentally
\cite{Schurig:2006:MEC,Farhat:2008:BCA,Liu:2009:BGP,Valentine:2009:OCD,Gabrielli:2009:COF}.

A completely different type of cloaking, which we call exterior cloaking
because the cloaking region is outside the cloaking device, was
introduced by Milton, Nicorovici, McPhedran and collaborators
\cite{Milton:2006:CEA,Nicorovici:2007:OCT,Milton:2008:SFG}.  They showed
that clusters of polarizable dipoles within a critical distance of a
flat or cylindrical superlens
\cite{Veselago:1967:ESS,Nicorovici:1994:ODP,Pendry:2000:NRM} are
cloaked. Anomalously localized fields generated by the interaction
between the induced dipoles and the superlens effectively cancel the
fields acting on the polarizable dipoles.  While larger objects do not
appear to be cloaked \cite{Bruno:2007:SCS}, Lai \etal 
\cite{Lai:2009:CMI} show that an object outside a superlens can be
cloaked if the appropriate ``antiobject'' is embedded in the superlens.

Ideally cloaking should be over a broad range of frequencies.  Most
cloaking methods are narrowband and approaches to obtain broadband
cloaking can have drawbacks, such as requiring frequency independent
relative dielectric constants or relative refractive indices less than
one
\cite{Li:2008:HUC,Leonhardt:2009:BIN,Liu:2009:BGP,Valentine:2009:OCD,Gabrielli:2009:COF}
which necessitate a surrounding medium with dielectric constant
sufficiently greater than one: thus such electromagnetic cloaking could
work underwater or in glass, but not in air or space. One proposal
without this drawback is the broadband interior cloaking scheme of
Miller \cite{Miller:2007:PC} which uses active cloaking controls rather
than passive materials. Here we also use active cloaking devices to
achieve broadband exterior cloaking. The principle is similar to that of
active sound control (see e.g \cite{Ffowcs:1984:RLA,Peterson:2007:ACS}), with the
fundamental novelty that we do not need a closed surface to suppress the
incident field in a region while not radiating significantly. Another
type of broadband exterior cloaking, using waveguides to guide waves
around a ``quiet zone'', has recently been introduced and confirmed
experimentally \cite{Smolyaninov:2009:AME}.

\section{Cloaking a single frequency}

For simplicity we just consider the two dimensional case, corresponding
to transverse electric or magnetic waves, so the governing equation is
the Helmholtz equation $\Delta u + k^2 u = 0$. Here $u(\Bx,\om)$ is the wave
field, $k = 2\pi/\la$ is the wavenumber and $\la = 2 \pi c_0/\om$ is the
wavelength at frequency $\om$ and at a constant propagation speed $c_0$.
We would like to cloak a region in the plane from a known probing
(incident) wave $u_i(\Bx,\om)$ supported in the frequency band $\om_0 +
[-B/2,B/2]$, where the central frequency is $\om_0$ and the bandwidth is
$B$.

The key to our cloaking method are devices that (a) cancel
the probing wave in the region to be cloaked and (b) radiate very little
waves away from the devices. To give a concrete example, let us take for
the cloaked region the disk $|\Bx|\leq \Ga$ and assume we measure the
radiation emitted by the devices on the circle $|\Bx| = \Gg > \Ga$. Thus
the device's field $u_d(\Bx,\om)$ must be so that (a') $u_d \approx
-u_i$ for $|\Bx| \leq \Ga$ and (b') $u_d \approx 0$ for $|\Bx| = \Gg$.

The devices can be idealized by $D$ points $\Bx_1,\ldots,\Bx_D$ with
$|\Bx_j| = \Gd$ and $\Ga<\Gd<\Gg$ so that the devices surround the
cloaked region. Because the device's field must solve Helmholtz equation and
become small far away, we take it as a linear combination of outgoing
waves emanating from the source points $\Bx_1,\ldots,\Bx_D$ with the
form \cite{Vasquez:2009:AEC}:
$$
 u_d(\Bx,\om) = \sum_{m=1}^D \sum_{n=-N}^N b_{m,n} H^{(1)}_n (k|\Bx -
 \Bx_m|) \exp[i n \theta_m ],
$$
where  $H^{(1)}_n$ is the $n-$th Hankel function
of the first kind and $\theta_m \equiv \arg(\Bx - \Bx_m)$ is the angle
between $\Bx - \Bx_m$ and $(1,0)$.  We seek coefficients $b_{m,n}$ so that
(a') holds on points of the circle $|\Bx| = \Ga$ and (b') holds on
points of the circle $|\Bx| = \Gg$. The control points are uniformly
distributed and at most $\la/2$ apart on each circle. The resulting
linear equations are solved in the least squares sense with the
Truncated Singular Value Decomposition in two steps. First we find
coefficients $b_{m,n}$ so that (a') holds, and second we find a
correction to enforce (b') while still satisfying (a'). (see
Appendix~\ref{app:coeff} for more details).
\section{Simulations}

\begin{figure}[t]
\begin{center}
\includegraphics[width=0.9\textwidth]{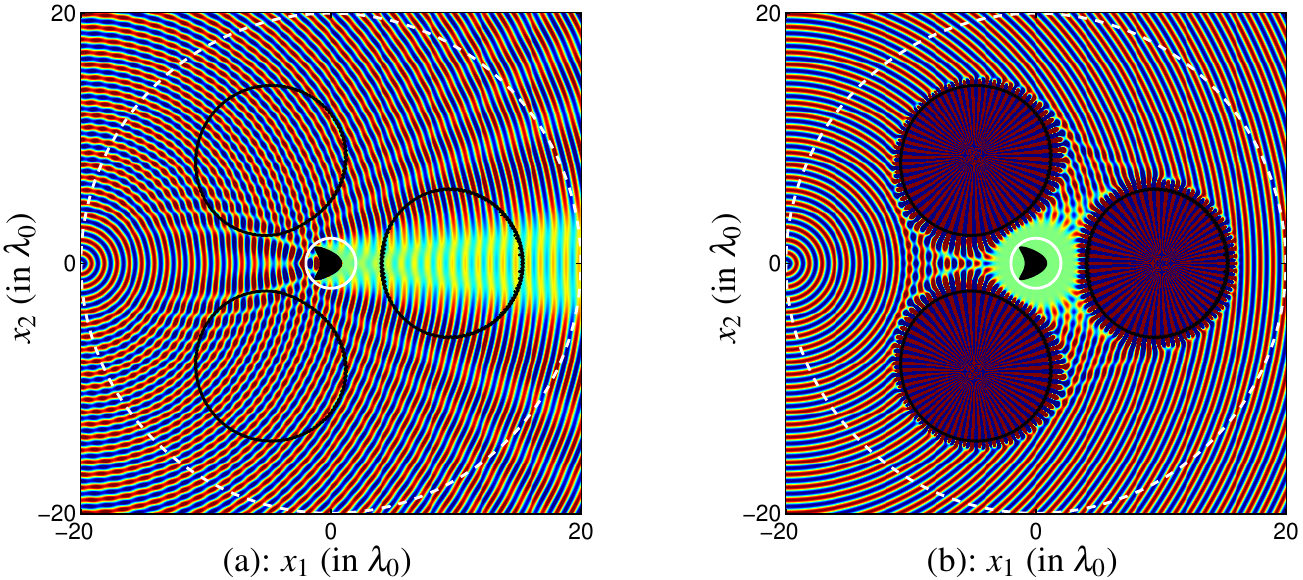}
\end{center}
\vspace{-1em}
\caption{Wave field at the central frequency $\omega_0$ when the cloaking
devices are (a) inactive and (b) active. Only the real part of the
fields is displayed, with a linear color scale going from $-1$ (dark
blue) to $1$ (dark red). All fields have been rescaled by
$|u_i(\Bx,\om)|$ to remove the geometric spreading of the point source.}
\label{fig:helm}
\end{figure}

We demonstrate cloaking in a regime that could correspond to Transverse
Electric microwaves in air (neglecting dispersion and attenuation),
where $u(\Bx,\om)$ is the transverse component of the electric field.
For the numerical experiments we took a
central frequency and bandwidth of $2.4$GHz, a propagation speed of $c_0
= 3\times 10^8$m/s and a central wavelength of $\la_0=12.5$cm.
Simulations suggest a minimum of three devices are needed to cloak
independently of the direction of the incoming waves.  In \figref{helm}
we show cloaking at the central frequency $\om_0/(2\pi) = 2.4$GHz of a
region of radius $\Ga = 2\la_0$ (solid white circle).  Here the devices
are located $\Gd = 10 \la_0$ from the origin and invisibility is
enforced at a distance $\Gg = 20\la_0$ from the origin (dashed white
circle). The incident wave is a point source 
originating at $\Bx_s = (-20,0)\la_0$ and modulated in frequency by a
Gaussian truncated to the bandwidth $\Hg(\om) = \sigma \sqrt{2\pi}
\exp(-\sigma^2 (\om - \om_0)^2/2)$, for $|\om - \om_0| \le B/2,$ and $0$
otherwise. We took $\sigma = 4/\om_0$. The scatterer is a perfectly
conducting ``kite'' obstacle \cite{Colton:1998:IAE} with homogeneous
Dirichlet boundary conditions, fitting inside the cloaked region. The
scattered field is computed using the boundary integral equation method
in \cite{Colton:1998:IAE}.

\begin{figure}[t]
\begin{center}
\includegraphics[width=0.55\textwidth]{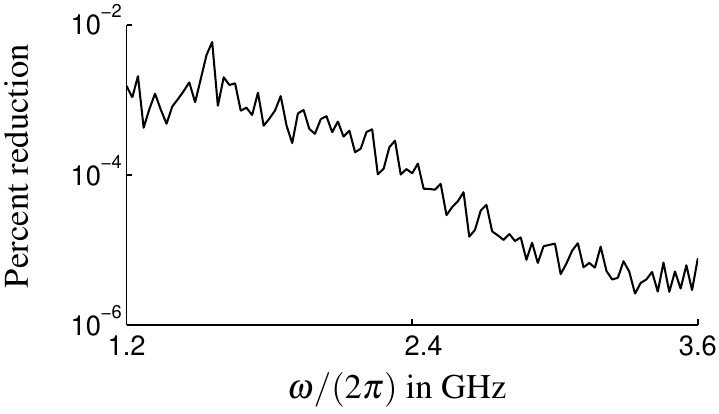}
\end{center}
\caption{Reduction of the scattering in percent achieved by the
cloaking devices over the bandwidth, measured in the $L^2$ norm on
$|\Bx| = 20\la_0$. The ordinates scale is logarithmic.}
\label{fig:perf}
\end{figure}

With the devices inactive (\figref{helm}a), the ``kite'' scatters the
incident field  and thus can be easily detected.  When
the devices are active (\figref{helm}b) they create a region with very
small fields while being nearly undetectable from far away. Since there
are almost no waves in the cloaked region, the scattered waves from the
object are greatly reduced, making the object invisible.
Quantitatively, the disturbance of the incident wave is $1.1\times
10^{-4}$\% of the field scattered without the devices, as measured on
the dashed white circle with the $L^2$ norm.  We carried the same
procedure for $N_{freq}=101$ frequencies in the bandwidth with similar results, as
can be seen in \figref{perf}.  Since the bandwidth is $100$\% of the central
frequency, broadband cloaking is possible with our approach.

\begin{figure}[t]
\begin{center}
\includegraphics[width=0.9\textwidth]{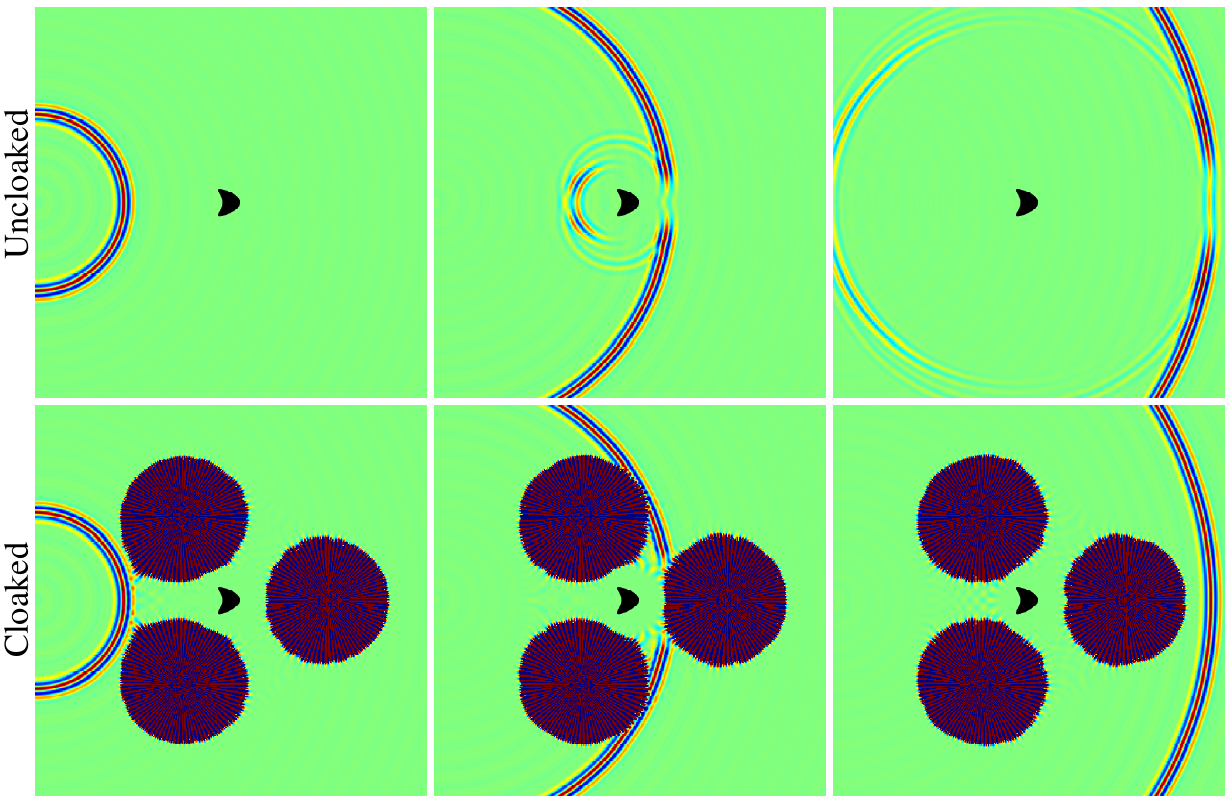}
\end{center}
\vspace{-1em}
\caption{Cloaking for a circular wave pulse. Top row: devices inactive
(\href{http://www.math.utah.edu/~fguevara/brd/}{Movie 1}). Bottom row: devices active (\href{http://www.math.utah.edu/~fguevara/brd/}{Movie 2}). The visualization
window in $\Bx$ is as in \figref{helm} and the scale is linear and relative to
the maximum amplitude on the plane of the incident field at each time.}
\label{fig:cloak}
\end{figure}

Several time snapshots (computed by taking the inverse Fourier transform
in time of the total fields) appear in \figref{cloak}. For full
animations see \figref{cloak} (\href{http://www.math.utah.edu/~fguevara/brd/}{Movie 1} and \href{http://www.math.utah.edu/~fguevara/brd/}{Movie 2}), which covers the time
interval $[0,T]$, with $T \approx 132$ns or the time
it takes for the wave to travel $50.5\la_0$. The devices make the
incoming wave disappear when it reaches the cloaked region, and then
rebuild the wave as it exits the cloaked region. This makes the object
virtually undetectable.

\begin{figure}[t]
 \begin{center}
 \includegraphics[width=0.95\textwidth]{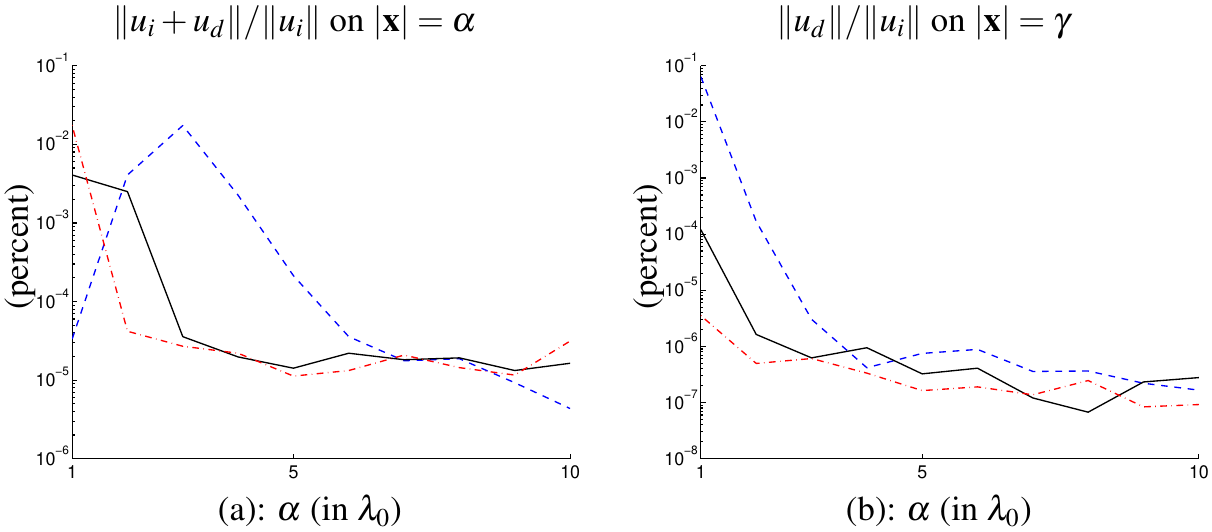}
 \end{center}
 \vspace{-1em}
 \caption{Cloak performance in terms of the cloaked region radius
 $\alpha$, as measured by (a) $\| u_i + u_d \|/\|u_i\|$ with the $L^2$
 norm on $|\Bx| = \alpha$ and (b) $\|u_d\|/\|u_i\|$ with
 the $L^2$ norm on $|\Bx| = \gamma$. Dashed, solid and dash-dotted
 lines correspond to $\om/(2\pi) = $ $1.2$GHz, $2.4$GHz and $3.6$GHz.}
 \label{fig:rcperf}
\end{figure}

Here the cloaked region is for visualization purposes deliberately small
($2\la_0=25$cm radius). However we have successfully cloaked regions up
to $10\la_0 = 1.25$m in radius as shown in \figref{rcperf}, where we
take
three devices with $|\Bx_m| = \delta= 5 \alpha$ and invisibility is
enforced on $|\Bx| = \gamma = 10 \alpha$. 

\begin{figure}[t]
 \begin{center}
  \includegraphics[width=0.45\textwidth]{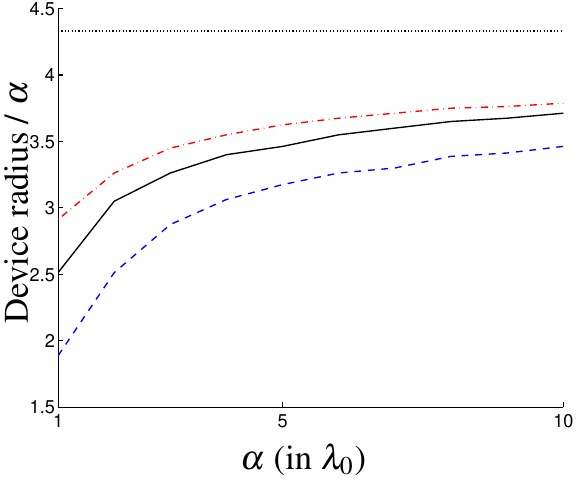}
 \end{center}
 \vspace{-1em}
 \caption{Estimate of the radius of circular cloaking devices relative
 to the radius of the cloaked region $\alpha$. Dashed, solid and
 dash-dotted lines correspond to $\om/(2\pi) = $ $1.2$GHz, $2.4$GHz and $3.6$GHz,
 respectively.  The device radius is estimated as the largest of the
 distances from a device point $\Bx_m$ to the level-set $\abs{u_d} = 100
 \max_{|\Bx|=\alpha} |u_i(\Bx,\om)|$.  If the devices were perfect
 circles, any device radius below the dotted line (at
 $\cos(\pi/6)\delta/\alpha = 5\sqrt{3}/2$) would indicate that the
 devices do not touch, i.e. they are three disjoint devices. }
 \label{fig:rcsz}
\end{figure}

The point-like devices could be problematic in practice because of the
$\CO(|\Bx-\Bx_m|^{-N})$ singularities near the devices. Fortunately the
point devices can be replaced  (with Green's identities) by
curves where the fields have reasonable amplitudes and where we can
control a single- and double-layer potential \cite{Colton:1998:IAE}.
These curves could be the circles suggested by the contours $\abs{u_d} =
100 \max_{|\Bx|=\alpha} |u_i(\Bx,\om)|$ (in black in \figref{helm}). The
radius of such devices for other cloaked region radii is estimated in
\figref{rcsz}. Since the devices do not completely surround the region
to be cloaked, exterior cloaking is possible at least for $\alpha
\leq 10\la_0$.  

A theoretical study of this cloaking method is ongoing, and we speculate
it generalizes to three dimensions and Maxwell equations.


\section*{Acknowledgments}
The authors are grateful for support from the National Science
Foundation through grant DMS-070978. An allocation of computer time from
the Center for High Performance Computing at the University of Utah is
gratefully acknowledged.

\appendix
\section{Finding the driving coefficients for the devices}
\label{app:coeff}
Let $\Bb$ be a vector with the $(2N+1)D$ possibly complex coefficients
$b_{m,n}$.  Denote by $\Bp_j^\Ga$ (resp.  $\Bp_j^\Gg$) the $N^\Ga$
(resp.  $N^\Gg$) control points on the circle $|\Bx| = \Ga$ (resp.
$|\Bx| = \Gg)$.  The coefficients $\Bb$, the number $N$ of terms in 
the expression for $u_d$ and the control points all
depend on the frequency $\om$.  Using the form for $u_d$ in the text,
construct a matrix $\BA$ of size $N^\Ga \times (2N+1)D$ and a matrix
$\BB$ of size $N^\Gg \times (2N+1)D$ such that $u_d(\Bp_j^\Ga,\om) =
(\BA \Bb)_j$ and $u_d(\Bp_j^\Gg,\om) = (\BB \Bb)_j$. We estimate the
driving coefficients as follows.
 {\bf Enforce (a')}: Use the Truncated Singular Value
 Decomposition (TSVD) to find coefficients $\Bb_0$ such that $\BA \Bb_0
 \approx - [ u_i(\Bp_j^\Ga,\om) ]$, in the least squares sense.
%
%
{\bf Enforce (b') while still satisfying (a'):} Find $\Bz$ as the
 solution to the least squares problem $ \BB (\Bb_0 + \BZ \Bz) \approx
 \Bzero$. Again the TSVD can be used for this step.  Here $\BZ$ is a
 matrix with columns spanning the nullspace of $\BA$ (a
 byproduct of the SVD in the previous step).

The coefficients are then $\Bb = \Bb_0 + \BZ \Bz$.  The heuristic for
$N$ in the numerical experiments is $N = \ceil{k (\delta - \alpha/2)}$,
where $\ceil{x}$ is the smallest integer larger than $x$.  The choice of
cut-off singular values can be used to control how well one wants to
satisfy (a') and (b').  We used a fixed $10^{-5}$ tolerance relative to
the maximum amplitude of $u_i$ on the control points.

\end{document}